\documentclass[jmp]{revtex4-1}
\usepackage{bm}
\usepackage{dcolumn}
\usepackage {amsfonts,amsmath}
\begin{document}

\title {The origin of holomorphic states in Landau levels from
non-commutative geometry, and a new formula for
their overlaps  on the torus.}

\author{ F. D. M. Haldane}
\email{haldane@princeton.edu} 
\affiliation{Department of Physics, Princeton
  University, Princeton NJ 08544-0708, USA}

\date{August 13, 2018}
\begin{abstract}
Holomorphic functions that characterize states in a
two-dimensional
Landau level been central  to key developments such as the Laughlin state.
Their origin has historically been attributed
to  a special property  of ``Schr\"{o}dinger
wavefunctions''  of  states in the ``lowest Landau level''.   It is
shown here that they instead  arise in
\textit{any} Landau level  as a generic mathematical property of the
Heisenberg description of the  non-commutative geometry of
guiding centers. When quasiperiodic boundary conditions are applied to
compactify the system on a torus, a new formula for the overlap
between holomorphic states, in the form of a discrete sum rather than
an integral, is obtained.
The new  formula is unexpected  
from  the previous  ``lowest-Landau level Schr\"{o}dinger
wavefunction'' interpretation.
 \end{abstract}
\maketitle

\section{introduction}

It is a well-known result that, in a uniform magnetic field, 
non-relativistic charge-$e$
electrons, moving on the flat two-dimensional Euclidean plane, and
treated in the ``symmetric gauge'', have lowest-Landau-level states
where the Schr\"{o}dinger wavefunctions have  the simple form
\begin{equation}
\psi(\bm r) \equiv 
\psi(x,y) \propto f(z)e^{-\frac{1}{4}z^*z/\ell^2}, \quad z(x,y) = x +
i y\, ,
\label{hol}
\end{equation}
where $2\pi \ell^2$ is the area of the 2D plane
through which one quantum $\Phi_0$ = $h/e$
of magnetic flux passes, and where $f(z)$ is a \textit{holomorphic
  function} (which follows from an orientation convention that makes
the sense of Landau orbits negative or clockwise).
This  remarkable appearance of holomorphic functions in a physical theory
has been at the heart of many important theoretical developments, such as
``conformal-block'' many-particle model wavefunctions, \textit{e.g.}, the
Laughlin state\cite{Laughlin}.

Despite the widespread belief that these
holomorphic structures are specific to  ``lowest-Landau level wavefunctions'',
it will here be shown that they are generic to any Landau level, and instead
are mathematical structures that derive from the non-commutative
geometry of  the ``guiding centers'' of Landau orbits. 
When this new viewpoint is applied to Landau levels compactified on
the torus by the application of quasiperiodic boundary conditions, a
key new formula for the overlap between two states emerges and is
presented here, whereby the 
continuum integral na\"{i}vely expected from the ``wavefunction''
interpretation  is replaced by a discrete sum.
The fact that the new formula presented here has apparently not
previously been reported may be a testament to the power of
misleading ``received wisdom'' to cause important aspects of a problem
to be overlooked or misinterpreted.

The conventional  ``Schr\"odinger wavefunction''  interpretation 
$\psi(\bm r)$ =  $\langle \bm r|\Psi\rangle$
naturally leads to the formula
\begin{equation}
\langle \psi_1|\psi_2\rangle = 
\int \frac{dx \wedge dy}{2\pi \ell^2} \, f_1(z)^*f_2(z)
e^{-\frac{1}{2}z^*z/\ell^2} \, .
\label{overlap}
\end{equation}
(With this normalization, $f(z)$ = 1 corresponds to a Gaussian
coherent state.)
However, the
interpretation of  (\ref{hol}) in terms of  ``Schr\"{o}dinger wavefunctions'',
while heuristically convenient,  does not provide  its fundamental meaning.

After projection into a Landau level by the quantization of the kinetic
energy,
the residual  degree of freedom
of a charged particle is
the   ``guiding center''  $\bm R$ of its  Landau orbit, the mean value of
the coordinate $\bm r$ averaged over an orbital period.
Its components obey the Heisenberg algebra of a non-commutative geometry
\begin{equation}
[R^x, R^y] = -i\ell^2\, .
\label{noncom}
\end{equation}
(The minus sign derives from the orientation 
 convention that makes $f(z)$ holomorphic rather than antiholomorphic.)
The uncertainty principle obeyed by the components of the
guiding-center coordinate removes the locality needed
for equivalence between the Schr\"{o}dinger and Heisenberg formulations
of quantum mechanics, invalidating the Schr\"{o}dinger description in
the projected space, and 
making only a (gauge-invariant) Heisenberg
description of their dynamics possible.       The holomorphic
structure is in fact  just 
 a generic consequence of the non-commutative geometry
(\ref{noncom}), \textit{unrelated} to any particular Landau-level
structure,
and  the holomorphic function $f(z)$ characterizes a Heisenberg state,
rather than a Schr\"{o}dinger wavefunction.

Furthermore,  the complex structure $z(x,y)$ which was hitherto imagined
to be given by $x + iy $, which in the ``lowest-Landau-level''
interpretation was inherited from the rotational
invariance (under the Euclidean metric) of  the dynamics of non-relativistic particles, 
is in fact an unspecified and arbitrarily-choosable parameter of
fundamental
representations of the Heisenberg algebra: 
\begin{equation}
a^{\dagger} = \frac{\bm e \cdot \bm R}{\surd 2\ell} \,  ,\quad 
a = \frac{\bm e^* \cdot \bm R}{\surd 2\ell} \, , \quad [a,a^{\dagger}] = 1
\, ,
\label{op}
\end{equation}
where $\bm e$ is a complex vector obeying
\begin{equation}
e^*_xe_y - e^*_ye_x = 2i\, .
\label{eee}
\end{equation}
Then
\begin{equation}
z(x,y) = \bm e \cdot \bm r \, .
\end{equation}

For technical reasons (to avoid complications from 
topologically-mandated edge states, and to maintain homogeneity) it is convenient to 
impose a (quasi)periodic boundary condition under a set of
translations 
$\{\bm L\}$ $\equiv$ $\mathbb L$
 that defines a lattice  with a unit cell of area $A$ = $2\pi
 N_{\Phi}\ell^2$ through which a flux $N_{\Phi} \Phi_0$
passes, where $N_{\Phi}$ is a positive integer.    This effectively
compactifies the
Euclidean plane to the torus, and leads to
``wavefunctions'' with the periodic  property
\begin{equation}
|\psi (\bm r + \bm L)|^2 = |\psi(\bm r)|^2\, .
\label{per}
\end{equation}
The natural generalization of (\ref{overlap}) is 
\begin{equation}
\langle \psi_1|\psi_2\rangle = 
\int_{\square} \frac{dx \wedge dy}{2\pi\ell^2} \, f_1(z)^*f_2(z)
e^{-\frac{1}{2}z^*z/\ell^2} \, ,
\label{periodicoverlap}
\end{equation}
where the integral is now over a unit cell of the lattice $\mathbb L$.

The central new mathematical  result presented here, which is ``unexpected'' from the
conventional Schr\"{o}dinger viewpoint, is that (\ref{periodicoverlap})
can be replaced by the \textit{finite sum}
\begin{equation}
\langle \psi_1|\psi_2\rangle = 
\frac{1}{N_{\Phi}}  {\sum_z}'
f_1(z)^*f_2(z) e^{-\frac{1}{2}z^*z/\ell^2}  \, ,
\label{sum}
\end{equation}
where the primed sum is over  a set of $(N_{\Phi})^2$ 
values $\{z\}$ = $\{\bm e\cdot \bm x\}$,  where the set $\{\bm x\}$ is chosen so that  
$N_{\Phi}\bm x$ $\in$ $\mathbb L$, and $ \bm x - \bm x'$ $\not \in
\mathbb L$ for $\bm x \ne
\bm x'$.   (This makes the $\bm x$ distinct modulo $\bm L$.)
Note that
\begin{equation}
\frac{1}{N_{\Phi}}{\sum_{z}}'\, 1 =
\int_{\square} \frac{dx \wedge dy}{2\pi\ell^2} \, 1
= N_{\Phi} .
\end{equation}

The expression (\ref{sum}) is valid for any choice of a set of
$(N_{\Phi})^2$ values of $\bm x$ that are distinct in the sense
described above.
In particular, if $(\bm L_1, \bm L_2)$ is a basis of the lattice, so
$\mathbb L$ = 
$\{m\bm L_1 + n\bm L_2, m,n \in \mathbb Z\}$, a possible choice for
$\{\bm x\}$ is
 a uniform grid in the unit cell,
\begin{equation}
\{\bm x\} = \left \{\frac{ (m\bm L_1 + n\bm L_2)}{N_{\Phi}}, m,n = 1,
\ldots N_{\Phi} \right\} \, ;
\end{equation}
however, the expression (\ref{sum}) is ``modular-invariant''
(independent of the choice of basis $(\bm L_1, \bm L_2)$).
The result may be viewed as a consequence of the underlying
non-commutative geometry.

 \section{Representation of states in a  Landau level}

When electrons move on a translationally-invariant 
two-dimensional (2D) surface with a uniform density of magnetic flux
passing through the surface, Landau quantization may occur, so that
their kinetic energy is quantized, and
the one-particle  spectrum consists of macroscopically-degenerate
Landau levels.             If the surface is flat, there is one
independent state in each Landau level
for each quantum  $\Phi_0$  of magnetic flux 
passing through the plane.
In the presence of the uniform magnetic field $\bm B$, the components of the
dynamical momentum
$\bm p$ = $-i\hbar \bm \nabla -e \bm A(\bm r)$ obey the Heisenberg algebra
\begin{equation}
[p_x,p_y]  =  i\hbar^2/\ell^2 \, .
\label{mom}
\end{equation}
The orientation of the plane is fixed by the
direction of the flux traversal: 
$\bm B \cdot \widehat {\bm n}$ = $\Phi_0/2\pi\ell^2$, where $\widehat {\bm n}$ is the
oriented unit normal of the surface.

 Semiclassically, this leads to energy-preserving motion of the 
momentum $\bm p$ along contours  of constant kinetic energy
$\varepsilon(\bm p)$ in momentum space, leading to periodic
(clockwise)
motion and
Bohr-Sommerfeld-Landau quantization if these contours are closed.
It will be assumed that the spectrum of $\varepsilon(\bm p)$ is
discrete, so that
\begin{equation}
\varepsilon(\bm p)|\psi_{n,\alpha} \rangle =
E_n|\psi_{n,\alpha}\rangle\,  ,\quad 
\langle \psi_{n\alpha} |\psi_{n\alpha'}\rangle =
                                                  \delta_{\alpha,\alpha'}
                                                  \, ,\quad
\langle \psi_{n\alpha} |\bm p|\psi_{n\alpha'}\rangle =
  \bm p_n\delta_{\alpha,\alpha'}\, ,
\end{equation}
where the degeneracy-label $\alpha$ counts one independent
state per Landau level per flux quantum passing through the plane,
\textit{i.e.},
 there are no additional degeneracies of a Landau level $E_n$.

The operator $\varepsilon (\bm p)$ is well-defined as a function of the
non-commuting components of $\bm p$
provided that the bivariate function $\varepsilon(\bm
p')$ of \textit{commuting} real components $\bm p'$ = $(p'_x,p'_y)$
(as when B = 0) has an
absolutely-convergent expansion in powers $ (p'_x)^m(p'_y)^n$.  To
construct
the quantum operator,  these
get replaced by  the symmetrized product of $m$ instances of the
operator $p_x$ and $n$ instances of $p_y$:  $(p_x')^m(p'_y)^n$  $\mapsto$
 $\{p_x,\ldots ,
p_x,p_y,\ldots p_y\}_{m+n}$, where the symmetrized product of $n$
operators
 is defined
so that $\{O,O,\ldots, O\}_n$ $\equiv$  $O^n$. 
 In practice, the  condition of absolute convergence of the expansion
will be satisfied if  $\varepsilon(\bm p)$ is given by a
 finite-degree bivariate polynomial.  At least semiclassically (and
as a plausible conjecture,   generally), the spectrum of
 $\varepsilon(\bm p)$ will be discrete (\textit{i.e.}, can contain Landau
 levels) in any open interval of energies
in which  the   set $\{\bm p';\varepsilon(\bm p') = E\}$ is compact 
(and will be empty in any  open interval where this set  is null).

The residual degree of freedom that gives rise to the macroscopic
degeneracy of Landau levels  is the guiding center $\bm R$ of the orbit,  where the
electron coordinate is decomposed in Landau level $n$ as
\begin{equation}
\bm r =  \bm R + \widehat {\bm n} \times (\bm p - \bm p_n) \ell^2/\hbar \, .
\end{equation}
The components of the guiding  center commute with the components of the 
momentum, and its components obey the Heisenberg algebra
(\ref{noncom})
which has the opposite chirality to that of the Heisenberg algebra
(\ref{mom})   of the components of the
momentum.

To form an orthonormal basis of states that span the Hilbert subspace
of the degenerate Landau level,   first choose a complex structure
defined by a complex vector $\bm e$ that satisfies (\ref{eee}).
Use (\ref{op}) to define a normalized guiding-center coherent state centered at the origin, with
a shape fixed by $\bm e$:
\begin{equation}
(\bm e^* \cdot\bm R)|0;\bm e\rangle = 0\, , \quad \langle 0;\bm
e|0;\bm e\rangle = 1\,.
\end{equation}
An orthonormal basis is then defined by
\begin{equation}
|m;\bm e \rangle =   \frac{1}{\surd m!} \left (\frac{\bm e\cdot \bm
    R}{\surd 2\ell}\right )^m |0, \bm e\rangle, \quad m = 0, 1,
2,\ldots.
\end{equation}
Now consider states
\begin{equation}
|f;\bm e\rangle = f(\bm e\cdot \bm R) |0; \bm e\rangle \, ,\quad
f(z) = \sum_{m=0}^{\infty} f^{(m)} z^m\, ,
\end{equation}
where the series is absolutely convergent, so
$f(z)$ is holomorphic.  
Then
\begin{equation}
\langle f_1;\bm e|f_2;\bm e\rangle = \sum_{m=0}^{\infty}
m!(2\ell^2)^m
f^{(m)*}_1f^{(m)}_2   =
\int_0^{\infty}\frac{ dx \wedge dy}{2\pi \ell^2} f_1(z)^*f_2(z) e^{-\frac{1}{2}z^*z/\ell^2}\, .
\end{equation}
Thus the structure (\ref{overlap})  that is commonly believed to be a
specific ``lowest  Landau level'' property has been reproduced for \textit{any} generic
Landau level, exposing its fundamental origin in  the non-commutative
geometry (\ref{noncom}) of the guiding centers.   In addition, the
complex structure $z(x,y)$ = $\bm e\cdot \bm r$ is revealed as a free
parameter, with an associated unimodular  (determinant 1) metric $g_{ab}$ =
$\frac{1}{2}(e^*_ae_b + e^*_be_a)$ that is not tied to the Euclidean
metric $\delta_{ab}$ that merely defines the Cartesian coordinate-system.

Note that  the \textit{same} state can have two different holomorphic
representations:
\begin{equation}
|\psi\rangle = |f;\bm e\rangle = |f';\bm e'\rangle
\end{equation}
If the state is held fixed while $\bm e$ is changed, the holomorphic
function
will in general change, so $f(z)$ has no meaning unless the complex
structure  $\bm e$ is also specified.   

\section{Quasiperiodic boundary conditions}

The unitary guiding-center translation operator that acts within the
degenerate manifold of states in a Landau level is
\begin{equation}
t(\bm d) = \exp(- i(d^xR^y-d^yR^x)/\ell^2) \, , \quad \bm d \in \mathbb
R_2\, , 
\end{equation}
with the action
\begin{equation}
t(\bm d) \bm R  = (\bm R +\bm  d) t(\bm d) \, .
\end{equation}
Note that
\begin{equation}
t(\bm d_1) t(\bm d_2) = \exp \left ({\textstyle\frac{1}{2}} i \varphi
  (\bm d_1, \bm d_2)\right ) t(\bm d_1
+ \bm d_2),
\quad \varphi(\bm d_1,\bm d_2) = (d_1^xd_2^{y} - d_1^yd_2^{x} )/\ell^2.
\end{equation}
When a complex structure is chosen,  with  $\bm e\cdot \bm R$ = $\surd
2 \ell a^{\dagger}$,
\begin {equation}
t(\bm d) = \exp \left ( (da - d^*a^{\dagger})/\surd 2\ell\right )
\, \quad d = \bm e \cdot \bm d  \, .
\end{equation}
This can usefully be normal-ordered as
\begin{equation}
t(\bm d) = e^{-\frac{1}{4}d^*d/\ell^2} 
e^{-d^*(a^{\dagger}/\surd  2\ell)}
e^{d(a/\surd 2\ell)} \, .
\end{equation}
Then
\begin{equation}
t(\bm d) |f,\bm e\rangle = |f_d,\bm e\rangle \, , \quad
f_d(z) = e^{-d^*(z + \frac{1}{2}d)/2\ell^2} f(z + d).
\end{equation}

The set $\{ t(\bm L), \bm L \in \mathbb L\}$  is a
mutually-commuting set that can be simultaneously diagonalized: for
all $\bm L $  in  $\mathbb L$,
\begin{equation}
t(\bm L)|\psi_{\alpha} (\bm K)\rangle = \xi(\bm L)^{N_{\Phi}} 
e^{i\bm K\cdot  \bm L} |\psi_{\alpha}(\bm K)\rangle, \quad \alpha = 1,\ldots , N_{\Phi},
\end{equation}
where $\xi(\bm L)$ is the \textit{parity} of $\bm L$: $\xi(\bm L)$ = 1
if $\frac{1}{2}\bm L$ $\in$ $\mathbb L$, and $-1$ otherwise.
Once the Bloch vector $\bm K$  that fixes the (quasi)periodic boundary
condition is chosen, there are $N_{\Phi}$
independent states in the Hilbert subspace.
The set  of $(N_{\Phi})^2$ translation operators $t(\bm L/N_{\Phi})$,
$\bm L/N_{\Phi}$ $\in$ $\{\bm x\}$ is a complete linearly-independent set of 
one-body operators compatible with the quasiperiodic boundary condition.

One can focus on the case $\bm K$ = 0, and obtain the general case as
\begin{equation}
|\psi_{\alpha}(\bm K)\rangle = e^{i\bm K\cdot \bm R}
|\psi_{\alpha}(\bm 0) \rangle.
\end{equation}
Since $\langle \psi_{\alpha}(\bm K)|\psi_{\alpha'}(\bm K)\rangle$ =
$\langle \psi_{\alpha}(\bm 0)|\psi_{\alpha'}(\bm 0\rangle$, it suffices
to establish (\ref{sum})  in the $\bm K$ = $\bm 0$ subspace, for the
result to have full generality.

Let $\Lambda$ = $\{\bm e\cdot \bm L\}$ be the mapping of the lattice
$\mathbb L$ to the complex plane, using the complex structure $\bm e$.
The $\bm K$ = $\bm 0$ holomophic quasiperiodic boundary condition is
\begin{equation}
f(z + L) = \xi(L)^{N_{\Phi}} e^{\frac{1}{2}L^*(z +  \frac{1}{2}L)/\ell^2} f(z).
\label{pbc}
\end{equation}
To solve this, it is useful to introduce\cite{modsig}
the ``modified sigma
function''
$\widetilde \sigma (z;\Lambda)$, related to the Weierstrass sigma 
function $\sigma(z;\Lambda)$ by
\begin{equation}
\widetilde \sigma (z;\Lambda) = e^{-\frac{1}{2}\gamma_2(\Lambda)z^2}\sigma
(z;\Lambda)  \, ;
\end{equation}
$\gamma_2(\Lambda)$ is a lattice invariant given by
\begin{equation}
\eta_i \equiv \zeta(\omega_i;\Lambda) = \gamma_2(\Lambda)\omega_i 
+ \frac{\pi\omega_i^*}{A(\Lambda)}\, , 
\end{equation}
where $\zeta(z;\Lambda)$ is the Weierstrass zeta function,
$A(\Lambda)$ is the area of the unit cell, and
$\omega_i$ is any primitive half-period of the lattice.
The modified sigma function (like the Weierstrass function) is odd and
holomorphic, with simple zeroes (only) at $z \in \Lambda$,
and  depends on the lattice $\Lambda$ without dependence on a choice
of basis (\textit{i.e.}, has ``modular invariance'').  Leaving the
dependence on $\Lambda$ implicit, 
It has the quasiperiodicity
\begin{equation}
\widetilde \sigma (z + L) = \xi(L)
e^{(\pi L^*/A)(z    + \frac{1}{2}L)}\widetilde \sigma (z).
\end{equation}
(Arguably\cite{modsig}, the modified function is the function
that Weierstrass \textit{should} have defined.)

The general solution of (\ref{pbc})  is
\begin{equation}
f(z) \propto e^{\frac{1}{2} L_0^*z/N_{\Phi}\ell^2}
\prod_{j=1}^{N_{\Phi}}\widetilde \sigma(z-w_j;\Lambda)\, ,\quad
\sum_{j=1}^{N_{\Phi}} w_j = L_0 \, ,
\label{wf}
\end{equation}
The value of $L_0\in \Lambda$ in $(\ref{wf})$ can be chosen for
convenience, and can be changed by a periodic redefinition of
the zeroes:
\begin{equation}
w_i \mapsto w_i + L_i, \quad L_0 \mapsto L_0 + \sum_i L_i \, ,
\end{equation} 
which merely modifies the (unspecified) normalization constant.
The ``wavefunction''
\begin{equation}
\psi(\bm r) =  f(z)e^{-\frac{1}{4} z^*z/\ell^2} = f(z) \left (
e^{-\frac{1}{2}\pi z^*z/A}\right )^{N_{\Phi}}
\end{equation}
has the property  (\ref{per}) that $|\psi(\bm r + \bm L)|^2$ = $|\psi(\bm r)|^2$.

It is now necessary to define an orthogonal basis of the $\bm K$ =
$\bm 0$ Hilbert subspace.   Let $(\bm L_1, \bm L_2)$  be a basis of
the lattice, with 
\begin{equation}
L_1^*L_2 - L_2^*L_1 = 2isA\, , \quad s =\pm 1,
\end{equation}
where $s$ is the orientation of the basis.
Then a basis $\{|\psi_k(\bm L_1,\bm L_2)\rangle,k = 1,\ldots, N_{\Phi}\}$ is defined by
\begin{equation}
 t({\textstyle\frac{\bm L_1}{N_{\Phi}}})|\psi_0(\bm L_1)\rangle =
                                                        -|\psi_0(\bm L_1)\rangle,\quad
|\psi_k(\bm L_1,\bm L_2)\rangle  =  
(-1)^k
t( {\textstyle\frac{k\bm  L_2}{N_{\Phi}}  })
|\psi_0(L_1)\rangle\, .
\end{equation}
Then $|\psi_k\rangle$ $\equiv$ $|\psi_k(\bm L_1,\bm L_2)\rangle$ =
$|\psi_{k+N_{\Phi}}\rangle$, and 
\begin{equation}
\langle \psi_k|\psi_{k'}\rangle = 0, \quad\mod (k-k',N_{\Phi}) \ne 0
\, .
\label{orthog}
\end{equation}

The holomorphic form  
$|f_0,\bm e\rangle$ = $|\psi_0 \rangle$  
is given by
\begin{equation}
f_0(z)  = C(L_1,\Lambda)
\prod_{j=1}^{N_{\Phi}} \widetilde \sigma (z-w_j),
\quad  w_j = ({\textstyle\frac{1}{2}} (N_{\Phi} + 1) - j)(L_1 /N_{\Phi}) \, .
\end{equation}
where $C(L_1,\Lambda)$ is a normalization constant.
It can also be represented as
\begin{equation}
f_0(z)   \propto  
\exp \left ({\textstyle\frac{1}{4}
    \frac{L_1^*}{L_1}\frac{z^2}{\ell^2}}\right )\chi_0(z)\, ,
\quad \chi_0(z) = 
\vartheta_*(u(z)|\tau)\, , \quad u(z) =
\frac{N_{\Phi}\pi z}{L_1}\, ,\quad \tau =
\frac{sN_{\Phi}L_2}{L_1} ,    
\end{equation}
where, for $N_{\phi}$ odd, $\vartheta_*(u|\tau)$ is the Jacobi theta function
$\vartheta_1(u|\tau)$ (with the classical definition that has
zeroes at $\{m\pi + n\pi \tau\}$), and is  $\vartheta_2(u|\tau)$ for even $N_{\Phi}$.

Since the absolute value of the normalization of the members of an
orthonormal basis set is an arbitrary choice, provided it is applied
equally to all states in the basis, the result (\ref{sum}) will be
established provided it can be  shown that it reproduces the
orthogonalities (\ref{orthog}).         
The full basis set (with a  common, but undetermined, normalization) 
has the holomorphic representation
\begin{align}
f_k(z) &= 
\exp \left ({\textstyle\frac{1}{4}  
\frac{L_1^*}{L_1}\frac{z^2}{\ell^2}}\right ) \chi_k(z),\\
\chi_k(z) &=
(-1)^k
\exp ({2i \mu_k ( u(z) + {\textstyle\frac{1}{2}}\mu_k\pi\tau)})
 \vartheta_*( u(z) + \mu_k \pi \tau |\tau)\, , \quad \mu_k = sk/N_{\Phi}
 \, .
\end{align}
This correctly has the property 
\begin{equation}
f_{k+N_{\Phi}} (z)= f_k(z) .
\end{equation}
It is also useful to note that the ``wavefunction''
$\psi_k(\bm r ;\bm L_1,\bm L_2)$ =  $f_k(z)\exp -\frac{1}{4}z^*z/\ell^2$ has the
properties
\begin{align}
\psi_k(\bm r + {\textstyle\frac{\bm L_1}{N_{\Phi}}}) &=  -\tilde \omega^{k}
e^{\frac{1}{2}\pi(L_1^*z-L_1z^*)/A} \psi_k(\bm r)\, ,\quad \tilde
                                                       \omega \equiv
                                                       e^{2\pi i
                                                       s/N_{\Phi}}  \,
  ,\\
\psi_k(\bm r + {\textstyle\frac{\bm L_2}{N_{\Phi}}}) &=  -
                                                       e^{\frac{1}{2}\pi(L_2^*z-L_2z^*)/A}
\psi_{k+1}(\bm r).
\label{no2}
\end{align}
Note these properties are independent of the choice of  complex
structure $\bm e$, and  the structure of the holomorphic representation
as a function of $z$
of these basis states is the same (up to a normalization constant) for
all choices of $\bm e$.

This must now be evaluated on the lattice
\begin{equation}
z \in \{z_{mn}\} \, ,  \quad z_{mn}= (mL_1 + nL_2)/N_{\Phi}\, ,
\quad u(z_{mn})  = m\pi   + n\pi \tau / N_{\Phi}\,. 
\end{equation}
First note that 
the relation(\ref{no2}) shows that  the normalizations are 
consistent (independent of $k$):
\begin{equation}
\frac{1}{N_{\Phi}}{\sum_{\bm x}}' |\Psi_k(\bm x)|^2  =
\frac{1}{N_{\Phi}}{\sum_{\bm x}}' |\Psi_0(\bm x)|^2 
\end{equation}
The lattice sum (\ref{sum}) for the overlap is
\begin{align}
\langle \psi_k|\psi_{k'} \rangle &=
\frac{1}{N_{\Phi}}
\sum_{m=1}^{N_{\Phi}}\sum_{n=1}^{N_{\Phi}}
F(z_{mn}, z^*_{mn}) \chi_k(z_{mn})^*\chi_{k'}(z_{mn}) \, \\
F(z,z^*) &= 
\exp \left ({\textstyle\frac{1}{4}  
\frac{L_1^*}{L_1}\frac{z^2}{\ell^2}}\right ) 
\exp \left ({\textstyle\frac{1}{4}  
\frac{L_1}{L_1^*}\frac{z^{*2}}{\ell^2}}\right ) 
\exp \left (-{\textstyle \frac{1}{2}\frac{z^*z}{\ell^2}}\right ) = 
\exp \left ({\textstyle\frac{1}{4}\frac{(L_1^*z -
  L_1z^*)^2}{L_1^*L_i\ell^2}}\right )\, , \nonumber \\
F(z_{mn},z^*_{mn}) &\equiv E(n) = 
\exp(-\pi |\tau -\tau^*|(n/4\pi N_{\Phi})^2)
\end{align}
so
\begin{equation}
\langle \psi_k|\psi_{k'} \rangle =
\sum_{n=1}^{N_{\Phi}} E(n)
\left (\frac{1}{N_{\Phi}}
\sum_{m=1}^{N_{\Phi}}
\chi_k(z_{mn})^*\chi_{k'}(z_{mn}) \right )\, .
\end{equation}
Then $\chi_{k}(z_{mn})$ factorizes as 
\begin{equation}
\chi_k(z_{mn}) =  (-1)^m \tilde \omega^{km} G_k(n)\, \quad
G_k(n) = (-1)^k\exp (2i\mu_k(n + {\textstyle\frac{1}{2}}sk)\pi\tau/N_{\Phi})
\vartheta_*((n + sk)\pi \tau/N_{\Phi}|\tau).
\end{equation}
The factorization of the $m$-dependence immediately gives the desired
orthogonality:
\begin{equation}
\langle \psi_k|\psi_{k'} \rangle =  \delta_{k,k'}
\frac{1}{N_{\Phi}}{\sum_{z}}' |f_0(z)|^2e^{-\frac{1}{2}z^*z/\ell_2}
\end{equation}
where 
\begin{equation}
\delta_{k,k'} \equiv \frac{1}{N_{\Phi} }\sum_{m=1}^{N_{\Phi}} \tilde \omega^ {m(k-k')},
\end{equation}
and the relation (\ref{sum}) is established.

\section{Discussion}

The state $|\psi_0(\bm L_1)\rangle $ was defined as a unique
eigenfunction of $t(\bm L_1/N_{\Phi}\rangle$, without reference to any
complex structure or metric,  which is why the structure of  the
zeroes of its
holomorphic representation  are independent of the choice of the
complex structure $\bm e$.    Only the normalization constant
$C(L_1,\Lambda)$ of $f_0(z)$  will vary as $\bm e$ is changed, and the
physical state described by its pattern of zeroes remains invariant.

Another unique state with   such a property is the antisymmetric state 
of $N$ = $N_{\Phi}$ fermions that completely fills the Landau level:
with a $\bm K$ = 0 boundary condition, this is 
\begin{equation}
F_0(z_1,\ldots , z_N)\propto  \widetilde \sigma(Z)\prod_{i<j}\widetilde \sigma
(z_i - z_j), \quad Z = \sum_i z_,
\end{equation}
Since this state is unique, it cannot depend on the choice of $\bm e$,
other than through its normalization.

Other model states such the  $\nu$ = $1/m$ Laughlin states\cite{Laughlin}
 do vary as the complex
structure is varied.       These states with $N_{\Phi}$ = $mN$,
$m > 1$ are the maximum-density states that belong to the kernel
of a  positive metric-dependent ``pseudopotential''  Hamiltonian\cite{vps}.
Let $\{\bm q\}$ be the reciprocal lattice of $\mathbb L$, the set where
$\exp i\bm q\cdot \bm L$ = 1 for all $\bm L$ $\in$ $\mathbb L$.  Also
let $g^{ab}$ be the inverse of a  unimodular Euclidean-signature
metric $g_{ab}$.   Then
\begin{align}
H(g) &= \sum_{m' < m} V_{m'} \sum_{i < j}P^g_{m'}(\bm R_i - \bm R_j), \quad
    V_{m'} > 0\, .\\ 
P^g_m(\bm R_1-\bm R_2)  &=   \frac{1}{N_{\phi}} \sum_{\bm q} 2L_m(u(\bm
       q))e^{-\frac{1}{2}u(\bm q) } e^{i\bm q\cdot (\bm R_1 - \bm
       R_2)}\, ,\quad
u(\bm q) = g^{ab}q_aq_b\ell^2\, ,
\end{align}
where $L_m(u)$ are Laguerre polynomials.  The Laughlin states on the 
torus\cite{fdmher} can be rewritten in the
recently-developed\cite{modsig}
modular-invariant 
holomorphic
representation as
\begin{align}
F(z_1,\ldots ,z_N)
 &= F_{\text{cm}}(Z) \prod_{i < j} \widetilde \sigma (z_i -
z_j)^m
\\
F_{\text{cm}}(Z) &\propto \prod_{j=1}^m\widetilde \sigma (Z-W_j)\, ,\quad
  \sum_j W_j = 0.
\end{align}
The free parameters $\{W_j\}$ parametrize the $m$-fold topological
degeneracy of the state.     The other parameter is the complex
structure $\bm e$.  In this case, the holomorphic state is only in the
kernel of  $H(g)$ if 
\begin{equation}
{\textstyle\frac{1}{2}}(e_a^*e_b + e_b^*e_a) = g_{ab}
\end{equation}
The Laughlin states  are a family of states continuously-parametrized
by a metric $g_{ab}$ which characterizes the  shape of the correlation
hole (or ``flux attachment'') surrounding each particle.  This is in stark contrast with the
filled Landau-level state, which is an uncorrelated Slater determinant
which does not vary with $\bm e$.

A long-standing technical problem has been how to perform the
transformation  to the
particle-hole conjugate of model many-fermion states in a partially-filled Landau
level, such as the Laughlin state.
The summation formula (\ref{sum}) simplifies this, at least in the
sense of reducing it to a finite algorithm in discrete mathematics.
The set of $N_{\Phi}$ particle coordinates in a filled Landau level can be split
up into $N_{\Phi}$ = $N + \widetilde N$ where the antiunitary
particle-hole transformation maps an $N$-particle state to a $\tilde
N$-particle state.     If $F(z_1,\ldots ,z_N)$ is
an antisymmetric many-particle holomorphic state, its (unnormalized)  particle-hole
conjugate
state is
\begin{equation}
\widetilde F(\tilde z_1,\ldots ,\tilde z_{\tilde N}) \propto
\frac{1}{(N_{\Phi})^N}{\sum_{z_1}}'\ldots {\sum_{z_N}}'
F(z_i,\ldots z_N)^*F_0(z_1,\ldots, z_N,\tilde z_1,\ldots ,\tilde
z_{\tilde N})\,.
\end{equation}
This is  the lattice sum version of the formal integral expression given by\cite{ph}.
While it involves a large sum over $(N_{\Phi})^{2N}$ terms, it is in
principle finite, as opposed to the formal integral of Ref. \cite{ph},
which in practice has never been carried out.

A second, perhaps more practical application of the lattice sum for
overlaps is that it indicates that complete information is contained in
a modular-invariant way in the $(N_{\Phi})^{2N}$ lattice configurations
$F(z_1,\ldots, z_N)$.  This  allows 
Metropolis Monte Carlo treatments 
of holomorphic model ``wavefunctions'' to be carried
out on a discrete grid on which the modified sigma function
has been tabulated.   This potentially leads to large speedups in such
calculations, and initial trials have been carried out\cite{jie}.

Finally, the existence of the formula (\ref{sum}) was quite unexpected,
at least to this author, who was  initially led to it by conjecture, followed by
numerical confirmation, and finally rigorous derivation.
The fact that it was apparently not previously found, is testament to
the misleading nature of the ``lowest-Landau-level wavefunction''
interpretation of the origin of holomorphic structures in Landau-level
physics, as opposed to the more-fundamental ``quantum-geometry''
explanation
presented here.

.
\begin{acknowledgements}
This work was supported by Department of Energy BES Grant DE-SC0002140.
\end{acknowledgements}

\end{document}